%% file: wshop97.tex
\documentstyle[11pt,epsf]{article}

\input{ws95defs}

\begin{document}

\title{Short - time scaling behavior of growing interfaces}
\author{M. Krech\\Fachbereich Physik, BUGH Wuppertal, 42097 Wuppertal,
Germany}
\date{}
\maketitle

\begin{abstract}
\input{absws97}
\end{abstract}

\section{Introduction}
 \input{inws97}

\section{Analytic Theory}
 \input{atws97}

\section{Monte-Carlo Results}
 \input{mcrws97}

\renewcommand{\thesection}{}
\section{Acknowledgment}
The author gratefully acknowledges partial financial support of this
work through the Heisenberg program of the Deutsche Forschungsgemeinschaft.

\input{bibws97}
\end{document}

%% file: ws95defs.tex
\newcommand{\be}{\begin{equation}}
\newcommand{\ee}{\end{equation}}
\newcommand{\ba}{\begin{array}}
\newcommand{\ea}{\end{array}}
\newcommand{\bt}{\begin{table}}
\newcommand{\et}{\end{table}}
\newcommand{\bes}{\begin{eqnarray}}
\newcommand{\ees}{\end{eqnarray}}
\newcommand{\bfig}{\begin{figure}}
\newcommand{\efig}{\end{figure}}

\newcommand{\Eq}[1]{Eq.(\ref{#1})}
\newcommand{\Eqs}[2]{Eqs.(\ref{#1}) and (\ref{#2})}

\renewcommand{\thesection}{\arabic{section}.}

%% file: absws97.tex
The short-time evolution of a growing interface is studied analytically
and numerically for the Kadar-Parisi-Zhang (KPZ) universality class.
The scaling behavior of response and correlation functions is
reminiscent of the ``initial slip'' behavior found in purely
dissipative critical relaxation (model A). Unlike model A the initial
slip exponent for the KPZ equation can be expressed by the dynamical
exponent $z$. In 2+1 dimensions $z$ is estimated from the short-time
evolution of the correlation function for ballistic deposition and for
the RSOS model.

%% file: inws97.tex
\setcounter{equation}{0}
Interface formation and growth are typical processes in nonequilibrium
systems. Two important examples are fluid flow in porous media
\cite{BarStan} and deposition of atoms during molecular beam
epitaxy (MBE) \cite{BarStan,SLKG}. It is expected that at times much
later than typical aggregation times and on macroscopic length scales
these interfaces develop a characteristic scaling behavior, where the
scaling exponents fall into certain dynamic {\em universality classes}
\cite{BarStan,SLKG,KS}. In certain cases, however,
interfaces can also show turbulent, i.e., spatial multiscaling
behavior \cite{Krug94}. Usually a $d$-dimensional interface is
embedded in $d+1$-dimensional space such that the interface position
at time $t$ can be described by a height function $h({\bf x},t)$,
where ${\bf x}$ denotes the lateral position in a $d$-dimensional
reference plane given by the surface of a substrate.
Complete information about the scaling behavior is contained in the
dynamic structure factor, which is related to the time displaced
height-height correlation function $C({\bf x}-{\bf x'},t,t') \equiv
\langle h({\bf x},t) h({\bf x'},t') \rangle - \langle h({\bf x},t)
\rangle \langle h({\bf x'},t') \rangle$, where a laterally
translational invariant system is assumed. For $t,t' \to \infty$ and
finite $|t-t'|$ the correlation function displays the asymptotic
scaling behavior
\begin{equation} \label{Cscal}
C({\bf x}-{\bf x'},t,t')=|{\bf x}-{\bf x'}|^{2\alpha}
F_C(|t-t'|/|{\bf x}-{\bf x'}|^z),
\end{equation}
where $\alpha$ denotes the {\em roughness} exponent and $z$ is the
{\em dynamic} exponent \cite{BarStan,SLKG}. For a laterally
translational invariant system the interfacial width $w^2(t) \equiv
\langle h^2({\bf x},t) \rangle - \langle h({\bf x},t) \rangle^2$ is
only a function of $t$ and displays the scaling behavior $w(t) \sim
t^\beta$ for late times, where $\beta = \alpha / z$ is the {\em
growth} exponent. For MBE as an example the scaling behavior displayed
in \Eq{Cscal} gives access to the exponents $\alpha$ and $z$ both
experimentally by reflection high energy electron diffraction (RHEED)
(see, e.g., chaper 16 of Ref.\cite{BarStan}) and by direct imaging using
a surface tunneling microscope \cite{KHHB} and theoretically by
continuum models \cite{BarStan,SLKG} and Monte-Carlo simulations
\cite{SLKG,PalLan}.

Continuum descriptions of interfacial growth processes can be obtained
from general symmetry principles and conservation laws obeyed by the
growth process \cite{BarStan}. For a wide class of growth processes
the resulting continuum model is given by the well-known
Kadar-Parisi-Zhang (KPZ) equation \cite{KPZ}, which reads
\begin{equation} \label{KPZeq}
\textstyle{\partial \over \partial t}h({\bf x},t)
= \nu \nabla^2 h({\bf x},t) + \textstyle{\lambda \over 2}
(\nabla h({\bf x},t))^2 + \eta({\bf x},t).
\end{equation}
The noise $\eta({\bf x},t)$ has a Gaussian distribution with
$\langle \eta({\bf x},t) \rangle = 0$ and
\begin{equation} \label{noise}
\langle \eta({\bf x},t) \eta({\bf x'},t') \rangle =
2D \delta ({\bf x}-{\bf x'}) \delta (t-t').
\end{equation}
The parameters $\nu$, $D$, and $\lambda$ are assumed to be constants
and averages $\langle \dots \rangle$ are taken over the noise
distribution.
In the long time limit \Eq{KPZeq} has a global symmetry which is
commonly denoted as Galileian invariance \cite{BarStan,KPZ}. An
important consequence is that the exponents $z$ and $\alpha$ of the
KPZ equation fulfill the scaling relation $\alpha + z = 2$. The
exponents of the KPZ equation are exactly known only in $d = 1$, where
$z = 3/2$ and $\alpha = 1/2$ due to the existence of a dissipation
fluctuation theorem \cite{FT,DH}. In $d = 2$ numerical investigations
indicate $z \simeq 1.6$ and $\alpha \simeq 0.4$ \cite{BarStan}. For $d
> 2$ the asymptotic scaling behavior is either governed by linear
theory ($\lambda = 0$, weak coupling regime) or by another set of
exponents inaccessible by analytical methods (strong coupling regime)
depending on the value of the effective coupling constant $g \equiv D
\lambda^2 / (4 \nu^3)$ \cite{BarStan,FT}. Furthermore, it is
interesting to note that the nonlinearity in \Eq{KPZeq} renders all
other possible nonlinearities irrelevant in the renormalization group
sense in the long-time limit. For intermediate times, however, the
presence of other nonlinearities in the growth equation gives rise to
various crossover phenomena \cite{BarStan,SKJJB}.

%% file: atws97.tex
\setcounter{equation}{0}
In Fourier space the KPZ equation \Eqs{KPZeq}{noise} is
equivalent to the dynamic functional ${\cal J}={\cal J}_0+{\cal J}_1$
\cite{FT,JSS,Lassig} which consists of the Gaussian part
\begin{equation} \label{Jhh0}
{\cal J}_0[\tilde{h},h]=\int {d^dq \over (2\pi)^d} \int_0^\infty dt
\left\{ D \tilde{h}({\bf q},t) \tilde{h}(-{\bf q},t)
- \tilde{h}({\bf q},t) \left({\partial \over \partial t}h(-{\bf q},t)
+ \nu {\bf q}^2 h(-{\bf q},t) \right) \right\}
\end{equation}
and the interaction part
\begin{equation} \label{Jhh1}
{\cal J}_1[\tilde{h},h]=-{\lambda \over 2} \int {d^dq_1 \over (2\pi)^d}
\int {d^dq_2 \over (2\pi)^d} \int_0^\infty dt\, {\bf q}_1 \cdot {\bf q}_2\,
\tilde{h}(-{\bf q}_1-{\bf q}_2,t) h({\bf q}_1,t) h({\bf q}_2,t) ,
\end{equation}
where $\tilde{h}({\bf q},t)$ is the Fourier transform of the response
field \cite{MSR}. The initial condition $h({\bf q},0)=0$, which is
implicitly assumed in \Eqs{Jhh0}{Jhh1}, breaks the temporal
translational invariance of the KPZ dynamics. The analytic treatment
of \Eqs{Jhh0}{Jhh1} is based on the identities
\begin{equation} \label{hh0}
\textstyle{\partial \over \partial t} h({\bf q},t = 0) = 2D
\tilde{h}({\bf q},t = 0) \quad \mbox{and} \quad
G({\bf q} = {\bf 0},t,t' < t) = 1,
\end{equation}
where the response function $G$ is given by the formal average $\langle
h(-{\bf q},t) \tilde{h}({\bf q},t') \rangle$ with respect to the dynamic
functional ${\cal J}[\tilde{h},h]$ \cite{MK97}. For details of the
field-theoretic treatment of \Eqs{Jhh0}{Jhh1} we refer to
Refs.\cite{JSS,MK97} and only quote the main results for later reference.

The correlation function $C$ and the response function $G$ are found
to obey the scaling relations
\begin{equation} \label{Cqtt}
C({\bf q},t,t' \ll t) = (t'/t)^{\theta}
|{\bf q}|^{-d-2\alpha} f_C(|{\bf q}|^z t) \quad \mbox{and} \quad
G({\bf q},t,t' \ll t) = (t'/t)^{\tilde{\theta}}
|{\bf q}|^{-d} f_G(|{\bf q}|^z t),
\end{equation}
respectively, where in contrast to model A \cite{JSS} the short-time
exponents $\theta$ and $\tilde{\theta}$ are given by
\begin{equation} \label{theta}
\theta = (d+4)/z - 2 = (d + 2 \alpha)/z \quad \mbox{and} \quad
\tilde{\theta} = 0
\end{equation}
at the nontrivial fixed point $(\lambda \neq 0)$ of \Eq{KPZeq}. In
$d=1$ the exact value $\theta = 4/3$ can be obtained from the exact
value $z = 3/2$ and is confirmed by a Monte-Carlo simulation of ballistic
deposition \cite{MK97}. From numerical estimates for $z$ in $d=2$ one
obtains $\theta \simeq 1.7$. The exponent relation given by \Eq{theta}
simply means that the short-time and the long-time scaling behavior of
the correlation function are {\em identical}, i.e., the short-time
scaling behavior can be obtained by extrapolating the $t'$-dependence
of $C({\bf q},t,t')$ from $t' \sim t$ to $t' = 0$. In fact, the
scaling relation given by \Eq{theta} can be derived independently by
analyzing the fluctuation spectrum of the interface displacement
velocity averaged over a macroscopic portion of the interfacial area
\cite{Krug91}. It should be noted, however, that the perturbative
analysis of Ref.\cite{MK97} only consitutes a rigorous proof of
\Eq{theta} for $d = 1$. For $d \geq 2$ one encounters the strong
coupling regime of \Eq{KPZeq} which can no longer be treated
analytically.

Finally, we remark that an alternative scaling form for $C$ can be
obtained from the definition of the growth exponent $\beta$ which
leads to $\theta = d/z + 2\beta$. The scaling behavior displayed in
\Eq{Cqtt} can then be written in the simplified form $C({\bf q},t,t'
\ll t) = t'^\theta g_C(|{\bf q}|^z t)$, where $g_C(y)=y^{-\theta} f_C(y)$.

%% file: mcrws97.tex
\setcounter{equation}{0}
The scaling behavior of $C({\bf q},t,t' \ll t)$ according to \Eq{Cqtt}
can be tested numerically by a Monte-Carlo simulation of simple
deposition models on lattices \cite{BarStan}. The continuum
description underlying \Eqs{KPZeq}{noise} is replaced by a discretized
description according to
\begin{equation} \label{hxthjn}
h({\bf x},t) = h({\bf x} = a {\bf j},t = n/(FL^d)) \equiv a\,h_{\bf j}(n),
\end{equation}
where the lattice constant $a$ is assumed to be the same both in
the plane of the substrate and perpendicular to it and ${\bf j} =
(j_1, \dots , j_d)$. The lattice has $L^d$ sites, $F$ is the incoming
particle flux, and $n$ is the number of deposited particles.
Furthermore, the incoming particle flux $F$ has been normalized to
unity, so that $t$ in \Eq{hxthjn} is dimensionless and given by the
number of deposited layers. Finally, $h_{\bf j}(n)$ defined by
\Eq{hxthjn} is also dimensionless and denotes the number of particles
deposited at lattice site ${\bf j}$ after $n$ particles have been
deposited on the lattice. Ballistic deposition on a two-dimensional
substrate is defined by the {\em deterministic} growth rule
\begin{equation} \label{bdgr}
h_{j,k}(n+1) = \mbox{max}(h_{j-1,k}(n), h_{j,k-1}(n), h_{j,k}(n)+1,
h_{j+1,k}(n), h_{j,k+1}(n)),
\end{equation}
(see, e.g., Ref.\cite{BarStan}), where the site $(j,k)$ in \Eq{bdgr}
has been selected randomly from the $L \times L$ sites of the lattice
and periodic boundary conditions are applied.

In order to measure the short time exponent given by \Eq{theta}
it is sufficient to probe the {\em integrated} time displaced
correlation function, i.e., one probes $C({\bf q} = {\bf 0},t,t')$
as described in Ref.\cite{MK97}.
\begin{figure}[t]
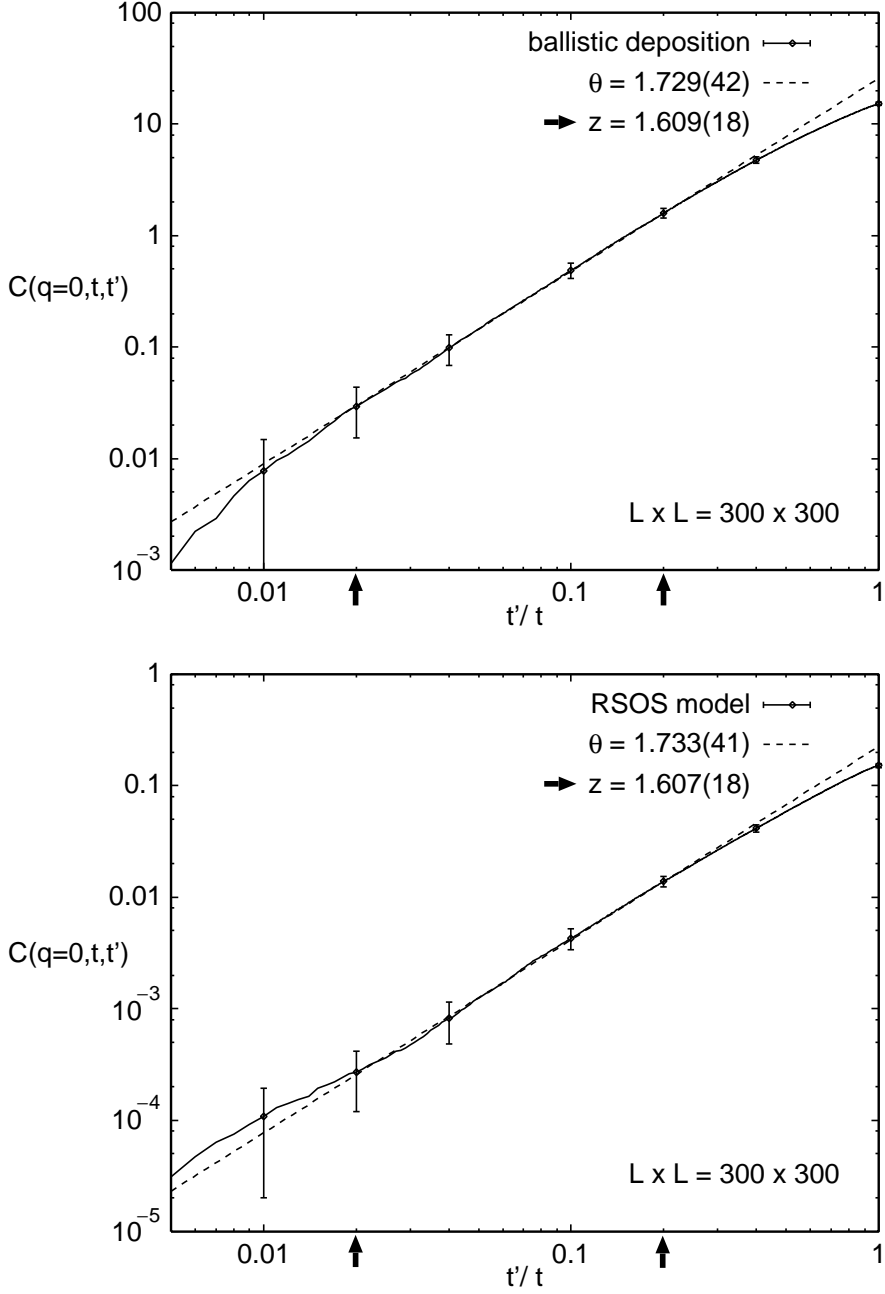

\centerline{\epsfbox{bd300.eps}}
\centerline{\epsfbox{rsos300.eps} \vspace*{2mm}}
\centerline{\parbox{14.5cm}{\caption{\protect\small{
Correlation function $C({\bf 0},t,t')$ in $d=2$ for ballistic
deposition (top) and the RSOS model (bottom) as a function
of $t'/t$ for $0.005 \leq t'/t \leq 1$ and $L\times L = 300 \times
300$ (solid line). Error bars are shown only at a few selected points
in time and represent one standard deviation. The dashed lines display
power laws with the measured short-time exponents $\theta = 1.729 \pm
0.042$ and $\theta = 1.733 \pm 0.041$, respectively. The data follow this
power law rather accurately in the interval $0.02 \leq t'/t \leq 0.2$
(bold arrows).}}
\label{CL0tt}}}
\end{figure}
Like a real deposition process the simulation is characterized by an a
priori unknown microscopic aggregation time $t_a$. The scaling behavior
of $C$ according to \Eq{Cqtt} can only be observed for $t' \gg t_a$.
On the other hand $t' \ll t$ is required for \Eq{Cqtt} to hold, so
that short-time scaling is restricted to the time window $t_a \ll t'
\ll t$. Furthermore, the lattice size $L$ must be chosen sufficiently
large in order to avoid the onset of finite-size crossover effects if
$t'^{1/z} \sim L$ when $t'$ is still much smaller than $t$. For the
simulation described here $t = 1000$ and $L \geq 200$ fulfill the
above requirements. In order to cope with the very small signal to
noise ratio in each measurement of $C({\bf 0},t,t')$ for $t' \ll t$
averages are taken over $2 \times 10^4$ realizations. These are
distributed over 20 individual runs at every point in time so that the
jackknife method can be applied for the data analysis. The result for
ballistic deposition according to \Eq{bdgr} and for the RSOS model in
$d = 2$ according to Ref.\cite{KK} is displayed in Fig.\ref{CL0tt},
where $C({\bf 0},t,t')$ is shown as a function of $t'/t$ for $t =
1000$ and $L = 300$. According to Fig.\ref{CL0tt} the interval $0.02
\leq t'/t \leq 0.2$ is available to determine the short-time exponent
$\theta$. From \Eq{theta} we obtain the estimate $z = 1.608 \pm
0.013$ by averaging over the two values for $\theta$ given in
Fig.\ref{CL0tt}. Finally, we note that according to the scaling
relation $\alpha + z = 2$ one has $\alpha = 0.392 \pm 0.013$ for the
roughness exponent. These values are in agreement with other numerical
data for $z$ and $\alpha$ in $d = 2$ (see chapter 8 of
Ref.\cite{BarStan} for a collection of recent estimates) and they
therefore provide some support for the general validity of \Eq{theta}.

%% file: wshop97.bbl
\begin{thebibliography}{99}
\bibitem{BarStan}
 A.-L. Barabasi and H.E. Stanley, {\em Fractal Concepts in Surface
 Growth} (Cambridge University press, New York, 1995) and references
 therein.
\bibitem{SLKG}
 S. Das Sarma, C.J. Lanczycki, R. Kotlyar, and S.V. Ghaisas, Phys.
 Rev. E {\bf 53}, 359 (1996) and references therein.
\bibitem{KS}
 J. Krug and H. Spohn, Phys. Rev. A {\bf 38}, 4271 (1988); J. Krug,
 J. Phys A {\bf 22}, L769 (1989); J. Krug, M. Plischke, and M.
 Siegert, Phys. Rev. Lett. {\bf 70}, 3271 (1993).
\bibitem{Krug94}
 J. Krug, Phys. Rev. Lett. {\bf 72}, 2907 (1994).
\bibitem{PalLan}
 S. Pal and D.P. Landau, Phys. Rev. B {\bf 49}, 10597 (1994) and
 references therein.
\bibitem{KHHB}
 J. Krim, I. Heyvaert, C. Van Haesendock, and Y. Bruynseraede, Phys.
 Rev. Lett. {\bf 70}, 57 (1993).
\bibitem{KPZ}
 M. Kadar, G. Parisi, and Y.-C. Zhang, Phys. Rev. Lett {\bf 56}, 889
 (1986).
\bibitem{FT}
 E. Frey and U.C. T\"auber, Phys. Rev. E {\bf 50}, 1024 (1994).
\bibitem{DH}
 U. Deker and F. Haake, Phys. Rev A {\bf 11}, 2043 (1975).
\bibitem{SKJJB}
 K. Sneppen, J. Krug, M.H. Jensen, C. Jayaprakash, and T. Bohr, Phys.
 Rev. A {\bf 46}, R7351 (1992).
\bibitem{JSS}
 H.K. Janssen, B. Schaub, and B. Schmittmann, Z. Phys. B {\bf 73}, 539
 (1989).
\bibitem{MSR}
 P.C. Martin, E.D. Siggia, and H.H. Rose, Phys. Rev. A {\bf 8}, 423
 (1973).
\bibitem{HH}
 P.C. Hohenberg and B.I. Halperin, Rev. Mod. Phys. {\bf 49}, 435 (1977).
\bibitem{Lassig}
 M. L\"assig, Nucl. Phys. {\bf B448}, 559, (1995).
\bibitem{MK97}
 M. Krech, Phys. Rev. E {\bf 55}, 668 (1997).
\bibitem{Krug91}
 J. Krug, Phys. Rev. A {\bf 44}, R801 (1991).
\bibitem{KK}
 J. M. Kim and J. M. Kosterlitz, Phys. Rev. Lett. {\bf 62}, 2289 (1989).
\end{thebibliography}
